\documentclass[]{spie}
\usepackage{amsmath,amsfonts,amssymb}
\usepackage{graphicx}
\usepackage[colorlinks=true, allcolors=blue]{hyperref}
\usepackage{booktabs}
\usepackage{float}
\usepackage{svg}
\usepackage{enumitem}

\title{FIRST-PL: Commissioning the first visible photonic lantern spectrograph for sub-diffraction-limit astronomy on Subaru/SCExAO}

\author[a,b,c]{S. Vievard}
\author[d]{E. Huby}
\author[d]{S. Lacour}
\author[c,e,f]{O. Guyon}
\author[g]{M. Lallement}
\author[d]{M. Nowak}
\author[h]{Y-J. Kim}
\author[b]{A. Walk}
\author[d]{J. Sarrazin}
\author[i]{S. Leon-Saval}
\author[i]{C. Betters}
\author[c,f]{J. Lozi}
\author[c,e]{M. Lucas}
\author[c,j]{V. Deo}
\author[k]{N.~Jovanovic}
\author[h]{M. Fitzgerald}
\author[i,l,m]{B. Norris}
\author[n]{T. Currie}
\author[c,f]{G. Singh}
\author[c,e]{Sandrine Juillard}
\author[d]{G. Perrin}

\affil[a]{Space Science and Engineering Initiative, College of Engineering, University of Hawai‘i, Hilo, HI 96720, USA}
\affil[b]{Institute for Astronomy, University of Hawaii at Manoa, Hilo, HI 96720 USA}
\affil[c]{Subaru Telescope, National Astronomical Observatory of Japan, Hilo, HI, USA}
\affil[d]{LESIA, Observatoire de Paris, Université PSL, Meudon, France}
\affil[e]{Steward Observatory, University of Arizona, Tucson, AZ 85721, USA}
\affil[f]{Astrobiology Center, Mitaka, Tokyo 181-8588, Japan}
\affil[g]{Univ. Grenoble Alpes, CNRS, IPAG, 38400 Saint-Martin-d'Hères, France}
\affil[h]{University of California, Los Angeles, CA 90095, USA}
\affil[i]{Sydney Institute for Astronomy, School of Physics, The University of Sydney, NSW 2006, Australia}
\affil[j]{Optical Sharpeners, Manosque, France}
\affil[k]{California Institute of Technology, 1200 E California Blvd, Pasadena, CA 91125, U.S.A.}
\affil[l]{Sydney Astrophotonic Instrumentation Laboratory, School of Physics, The University of Sydney, Sydney, NSW 2006, Australia}
\affil[m]{Astralis, School of Physics, University of Sydney, Sydney 2006, Australia}
\affil[n]{Department of Physics and Astronomy, University of Texas, San Antonio, TX, USA}

\begin{document}

\maketitle

\begin{abstract}
FIRST-PL (Fibered Imager foR a Single Telescope - Photonic Lantern) is a newly commissioned visible-light instrument on Subaru/SCExAO achieving spectroscopy below the diffraction limit. The instrument uses a Photonic Lantern (PL)—converting multimode fiber into 19 single-mode outputs—feeding a mid-resolution spectrograph (R~3000, 630-790 nm). On-sky performance demonstrates 40\% injection efficiency at 680 nm (Strehl ~30\%) and 12$\times$ throughput improvement over single-mode fibers. Three operational modes enable spectro-astrometry (50 $\mu$as precision demonstrated on $\beta$-CMi), image reconstruction, and high-contrast imaging. FIRST-PL represents a significant advancement in high-throughput photonic instrumentation.
\end{abstract}


\section{INTRODUCTION}
\label{sec:intro}

High-resolution spectroscopy is essential for advancing our understanding of the Universe, playing a pivotal role in fields ranging from stellar physics to exoplanet detection and characterization. In exoplanetary science, high-resolution observations make it possible to identify companions with narrow emission features, resolve fine kinematic structures, and isolate weak planetary signals via Doppler shifts—ultimately yielding precise measurements of orbital velocities, masses, and atmospheric compositions~\cite{snellen10, brogi12, currie2023ppvii}. Furthermore, high-dispersion spectra directly constrain the spin-orbit obliquities and rotation periods of both host stars and exoplanets~\cite{bryan_constraints_2017,bryan_obliquity_2020}, offering crucial insights into their formation pathways and dynamical evolution~\cite{lissauer1993planet}.

SMF-fed spectrographs provide several key advantages for high-resolution spectroscopy. First, they ensure optimal (minimal) background contamination, maximizing sensitivity in background-limited regimes such as faint-target observations (suppressing airglow, zodiacal, and thermal light) or high-contrast exoplanet spectroscopy (mitigating the speckle halo and exozodiacal light). Second, by spatial filtering to reject wavefront distortions, SMF feeds enhance spectral line profiles and enable highly compact, ultra-stable spectrograph architectures. A growing suit of SMF-based instruments are already operational or undergoing commissioning across major facilities, including Subaru/REACH~\cite{kotani2020reach}, Keck/KPIC~\cite{delorme2021keck}, Subaru/GLINT~\cite{martinod2021scalable}, Subaru/FIRST~\cite{vievard2023single}, VLT/HiRISE~\cite{vigan2024first}, and Keck/HISPEC~\cite{mawet2024fiber}. Despite these strengths, SMF coupling efficiency degrades significantly under optical aberrations—particularly in the visible regime—limiting most current applications to bright targets in the near-infrared.

A persistent challenge in observational astrophysics is that critical spatial and spectral structures often reside at angular scales well below the diffraction limit of even the largest modern telescopes. Although standard imaging methods struggle to probe features finer than $\lambda/D$, sub-diffraction signals can theoretically be recovered under high signal-to-noise conditions. The classic diffraction limit and Rayleigh criterion are not fundamental physical barriers to measuring tiny angular scales; rather, they reflect the observational difficulty of disentangling optical aberrations from intrinsic source structure. This distinction is central to spectroastrometry: a stellar photocenter shift on the order of tens of microarcseconds is, in principle, measurable to extreme precision in a photon-rich, well-calibrated instrument, despite being orders of magnitude smaller than $\lambda/D$. In practice, achieving this limit is constrained not by diffraction itself, but by the systemic challenge of separating true photocenter displacements from wavefront and instrumental drifts—the exact ambiguity that hinders sub-diffraction-limit imaging in general.

Sparse Aperture Masking (SAM)\cite{haniff1987first} represented a major milestone in overcoming this obstacle. By placing a non-redundant mask in the pupil plane, SAM divides the aperture into discrete sub-pupils and coherently recombines their light in the focal plane to effectively isolate wavefront errors from object features. However, SAM suffers from poor overall throughput, low contrast limits~\cite{gauchet2016sparse}, and an inability to support high-resolution spectroscopy. To address these drawbacks, fibered pupil remapping was introduced~\cite{perrin_high_2006}. By employing SMFs to spatially filter and route light from individual subapertures prior to interference, pupil remapping improves both fringe stability and pupil coverage. Crucially, using single-mode fibers allows the recombined light to directly feed a spectrograph, unlocking high-resolution spectroscopy at and below the telescope's diffraction limit~\cite{huby_first_2013,vievard2023single}.

Building upon these principles and the initial on-sky validation of a visible photonic lantern (PL) at the Subaru Telescope~\footnote{The authors wish to acknowledge how fortunate they are to perform observations from Maunakea, and acknowledge the indigenous Hawaiian communities who have had, and continue to have, \textit{Kuleana} (translated into English as responsibility) to this land.}~\cite{vievard2023single}, we present the commissioning and opto-mechanical upgrades of FIRST-PL. Installed on the Subaru Coronagraphic Extreme Adaptive Optics~\cite{jovanovic_subaru_2015} (SCExAO) instrument, FIRST-PL builds directly upon this initial demonstration, incorporating key hardware upgrades to transition the instrument into a commissioned instrument. In this paper, we describe the revised instrument architecture, the offered observing modes, we discuss data analysis and show early on-sky commissioning results.

\section{INSTRUMENT DESIGN AND ARCHITECTURE}

\subsection{Photonic Lantern Principle}
\label{subsec:pl_principle}

Our core objective is to deliver a high-throughput spectro-imaging solution that combines the key advantages of single-mode fiber (SMF) spectroscopy—optimal background rejection, exceptional spatial filtering, precision, stability, and compactness—with the ability to probe spatial and spectral structure well below the telescope diffraction limit. 

\begin{figure}[htbp]
\centering
\includegraphics[width=0.5\textwidth]{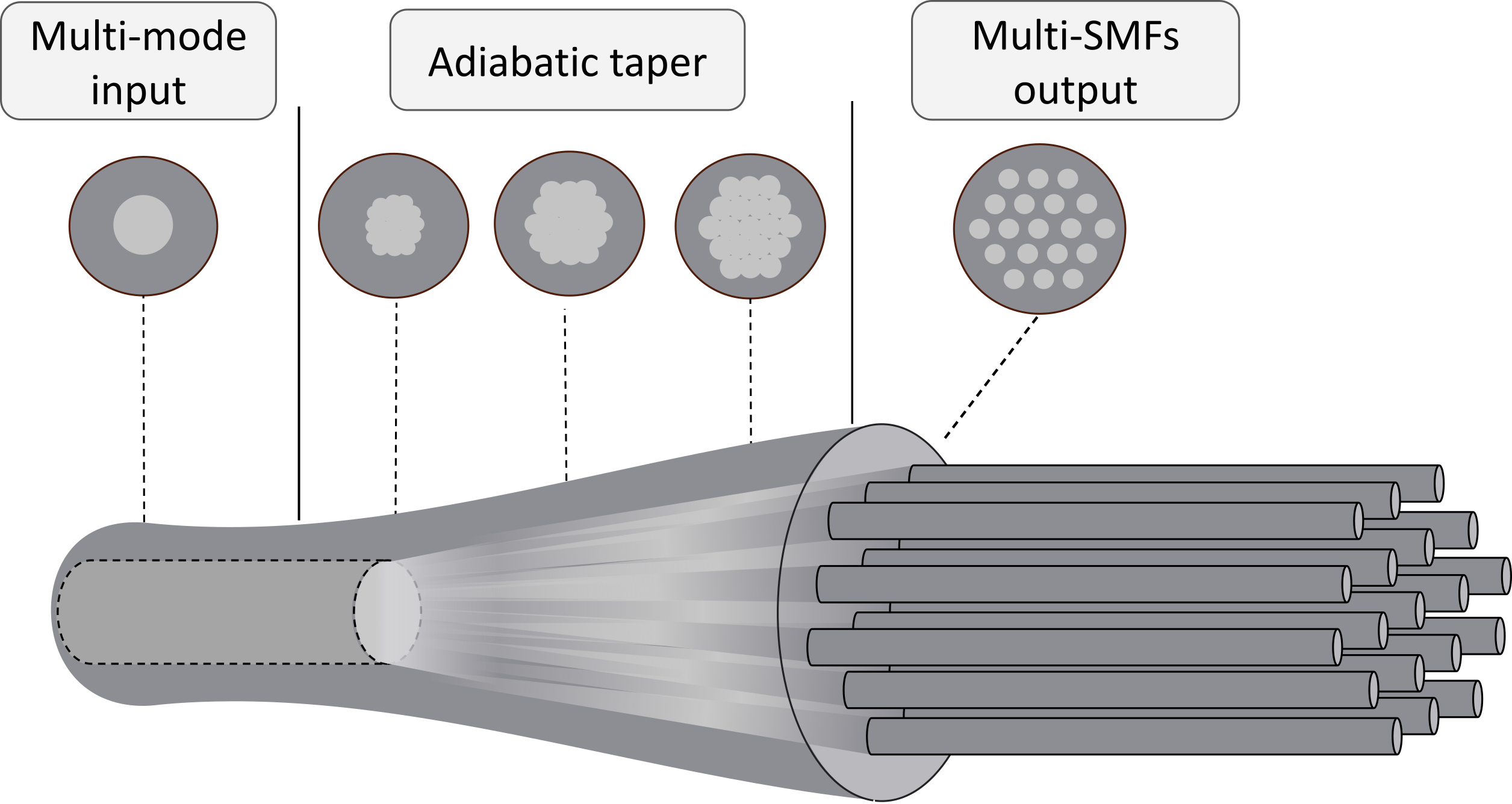}
\caption{Schematic representation of a 19-port photonic lantern~\cite{vievard_spectroscopy_2024}. Light incident on the multimode (MM) input waveguide undergoes a low-loss adiabatic transition through a tapered region, efficiently mapping the incoming spatial modes into an array of uncoupled single-mode fiber (SMF) outputs.}
\label{fig:pl}
\end{figure}

This is achieved via a photonic lantern (PL), an optical fiber device that adiabatically (i.e., slowly and losslessly) transitions a multimode (MM) input into an array of SMFs (Figure~\ref{fig:pl}). Within the PL, the spatial information present at the MM input is efficiently decomposed into orthogonal spatial modes, each corresponding ideally to a specific SMF output core. This architecture offers several distinct advantages:

\begin{itemize}[nosep]
    \item \textbf{High efficiency with photonic lanterns:} The adiabatic transition between the MM and SM regimes ensures high throughput, greater than $90\%$~\cite{Birks:15}. Thanks to the MM input, most of the source’s light is coupled in the PL --- this is critical at visible wavelengths, where coupling light in a single SM core would be challenging, requiring ~0.001” alignment of the source to the fiber~\cite{El_Morsy_2022} and high Strehl ratio (very difficult to achieve in visible light). Instead, the MM input has a ~0.1” core diameter, making the coupling efficiency much more robust to misalignment and wavefront errors. The robustness and high throughput of MM has long been established~\cite{jovanovic_efficient_2017}, and adopted for several instruments (eg. Subaru Prime Focus Spectrograph~\cite{tamura2016prime}, PFS).
    \item \textbf{Sub-Diffraction Spatial Encoding:} Spatial distribution and source offsets at the MM input are directly encoded into the relative output flux distribution among the SMFs~\cite{kim_potential_2024}.
    \item \textbf{Narrow-Field Integral Field Unit (IFU):} Dispersing the SMF output pigtails via a downstream spectrograph effectively transforms the PL into a compact, low-noise, full-efficiency narrow-field IFU.
    \item \textbf{Wavefront Sensing \& Self-Calibration:} By encoding input spatial phase and amplitude variations into the relative intensities of the SM outputs, the PL acts as a co-located focal-plane wavefront sensor~\cite{norris2020all}. This breaks phase ambiguities inherent to conventional intensity imaging (e.g., distinguishing signs of astigmatism~\cite{norris_all-photonic_2020}), allowing FIRST-PL to disentangle optical aberrations from true astrophysical structures for passive self-calibration or active wavefront control.
\end{itemize}

\subsection{Integration on SCExAO}
\label{subsec:scexao}

\begin{figure}[htbp]
\centering
\includegraphics[width=\textwidth]{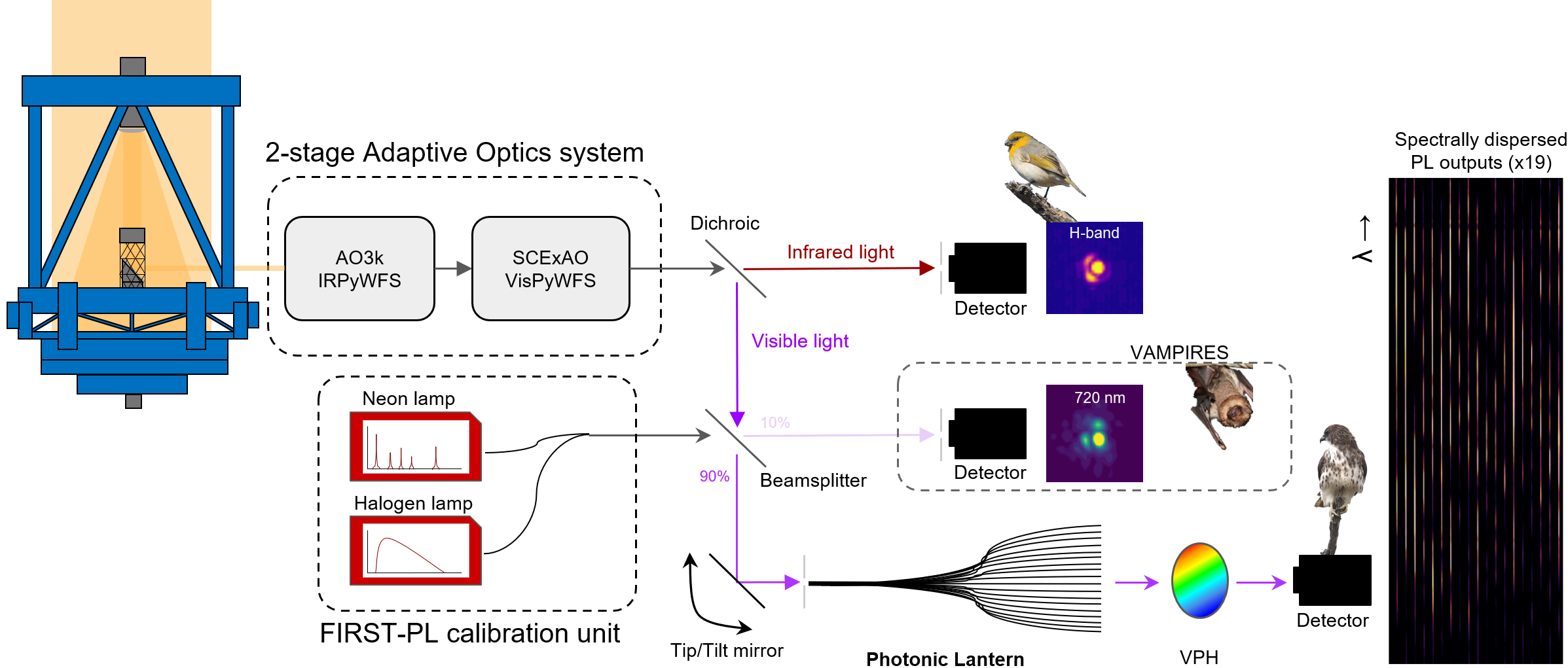}
\caption{System architecture and beam-path layout of FIRST-PL on SCExAO. Light corrected by the two-stage AO system (AO3k + SCExAO) is split by a dichroic between the NIR and visible paths. The visible light is split by a 90/10 beamsplitter: 90\% of the light is directed to FIRST-PL while 10\% is directed to VAMPIRES, providing focal plane image. A piezo tip-tilt mirror is used to adjust the injection into the PL. The PL outputs are dispersed onto a detector ($R \sim 3\,000$). Simultaneous focal-plane telemetry across three channels provides comprehensive diagnostic capability during observations. A Neon lamp and a Halogen lamp can be inserted in the optical path for calibration purposes. }
\label{fig:architecture}
\end{figure}

FIRST-PL is fully integrated downstream of the two-stage adaptive optics system at the Subaru Telescope, composed of AO3k~\cite{lozi2024ao3k} (feeding the IR Pyramid Wavefront Sensor, IRPyWFS) and SCExAO (feeding the Visible Pyramid Wavefront Sensor~\cite{Lozi_2019}, VisPyWFS), as illustrated in Figure~\ref{fig:architecture}. 

The optical path and calibration workflow are organized as follows:

\begin{itemize}[nosep]
    \item \textbf{Dichroic and Beam Splitting:} The corrected light is split by a dichroic filter, transmitting infrared light to the NIR focal-plane imaging channel while reflecting visible light to visible instruments. A second beam splitter routes $10\%$ of the visible light toward the VAMPIRES~\cite{lucas2024visible} instrument for simultaneous visible focal-plane imaging ($720\text{~nm}$ or multi-band images~\cite{lucas2024visible}), sending the remaining $90\%$ toward FIRST-PL.
    \item \textbf{FIRST-PL injection unit (see Figure~\ref{fig:injection_stage}):} A newly installed piezo-driven tip/tilt (TT) mirror is integrated directly upstream of the PL focal plane. This TT mirror serves a dual purpose: precise fine-alignment to maximize injection efficiency into the PL, and active modulation of the beam position during science exposures. Dedicated hardware electronics were implemented to hardware-trigger and synchronize the TT mirror movements with the detector acquisition. The PL is also mounted on a fine 3-axis translation stage assembly, used for alignment as well.
    \begin{figure}[htbp]
    \centering
    \includegraphics[width=\textwidth]{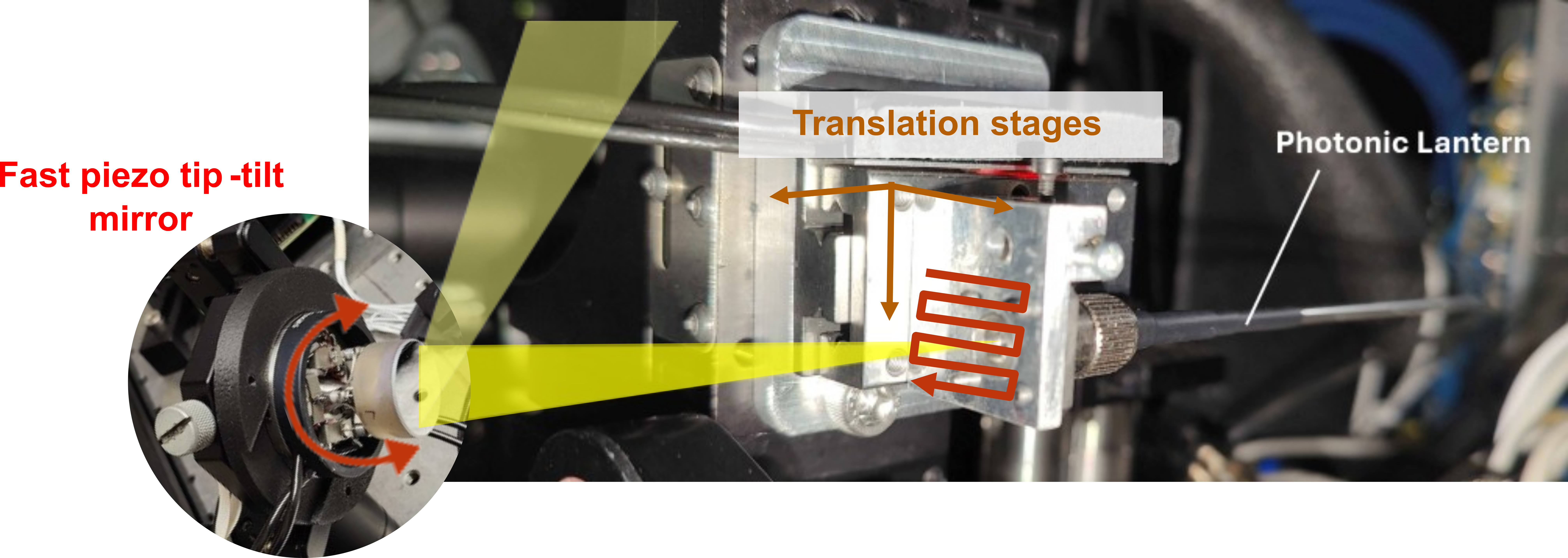}
    \caption{Opto-mechanical layout of the FIRST-PL injection stage. A newly installed fast piezo tip-tilt mirror modulates the incident beam onto the input face of the photonic lantern, which is mounted on high-precision translation stages. The red raster path illustrates the dynamic beam scanning pattern used across the lantern’s input field during calibration and observation.}
    \label{fig:injection_stage}
    \end{figure}
    \item \textbf{FIRST-PL Calibration Unit:} Positioned upstream, a dedicated calibration unit featuring Neon and Halogen lamps provides accurate wavelength calibration and flat-fielding capabilities prior to science observations.
    \item \textbf{Spectrograph:} Light emerging from the 19 SMF output pigtails is arranged in a linear V-groove array and collimated by an objective lens. Downstream, a Volume Phase Holographic (VPH) grating disperses the light onto the science camera through an imaging lens system (Figure~\ref{fig:spectrograph_modes}, top). To support polarimetric differential imaging (PDI), a Wollaston prism is mounted on a motorized stage immediately following the objective:
    \begin{itemize}[nosep]
        \item \textbf{Wollaston OUT (Standard Spectroscopy Mode):} The unpolarized beam yields 19 individual dispersed spectra across the detector (Figure~\ref{fig:spectrograph_modes}, bottom-left). On-sky spectra clearly show telluric oxygen absorption features as well as localized astrophysical features such as H$\alpha$ emission.
        \item \textbf{Wollaston IN (Polarimetric Mode):} The prism splits the output of each SMF core into orthogonal linear polarization states, doubling the spatial traces on the detector to 38 individual spectra (Figure~\ref{fig:spectrograph_modes}, bottom-right) for simultaneous polarimetric measurements.
    \end{itemize}
    \begin{figure*}[htbp]
    \centering
    \includegraphics[width=\textwidth]{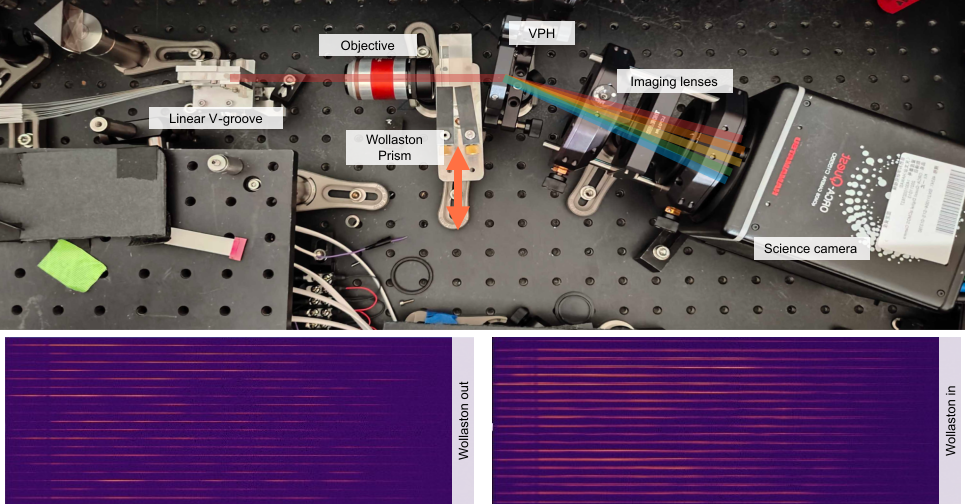}
    \caption{\textbf{Top:} Opto-mechanical layout of the FIRST-PL spectrograph assembly, showing the linear V-groove array output, objective collimator, motorized Wollaston prism, VPH grating, camera imaging lenses, and the science detector. \textbf{Bottom-left:} On-sky raw frame obtained with the Wollaston prism removed (OUT mode), displaying 19 individual spatial spectra with visible atmospheric oxygen absorption bands and localized H$\alpha$ emission. \textbf{Bottom-right:} On-sky raw frame on a distinct target with the Wollaston prism inserted (IN mode), splitting the light into 38 spectra for polarimetric differential imaging.}
    \label{fig:spectrograph_modes}
\end{figure*}
\end{itemize}

This multi-instrument feed provides three simultaneous telemetry streams during observations: (1) dispersed $19$-port PL spectra, (2) visible focal-plane imaging via VAMPIRES, and (3) NIR focal-plane imaging, enabling real-time calibration and diagnostic capabilities.

A summary of the technical parameters of the FIRST-PL instrument is provided in Table~\ref{tab:first_pl_specs}.

\begin{table}[htbp]
\centering
\caption{Key technical parameters and operational capabilities of FIRST-PL.}
\label{tab:first_pl_specs}
\begin{tabular}{lll}
\toprule
\textbf{Parameter} & \textbf{Value} & \textbf{Notes / Description} \\
\midrule
Operating Wavelength & $620 - 780\text{ nm}$ & Visible band coverage \\
Spectral Resolution  & $R \sim 3\,000$       & Enabled by VPH grating dispersion \\
Field of View (FoV)  & $80\text{ mas @ } f/8$ & Defined where injection drops to $50\%$ of peak center value~\cite{vievard_spectroscopy_2024} \\
Exposure Times       & $7.2\,\mu\text{s} - 1800\text{ s}$ & Supports fast or slow readout modes \\
\bottomrule
\end{tabular}
\end{table}

\section{OBSERVING MODES AND CALIBRATION STRATEGIES}
\label{sec:observing_modes}

FIRST-PL supports two observation modes depending on the science application and target topology: Spectro-Astrometry and Image Reconstruction (further categorized into On-Axis and Off-Axis configurations). These operational modes determine the data acquisition protocol, the dynamic control of the fast TT mirror, and the associated calibration overheads.

\subsection{Spectro-Astrometry Mode}
\label{subsec:spectro_astrometry}

Spectro-astrometry aims to extract sub-diffraction spatial information ($\ll \lambda/D$) and precise photocenter displacements as a function of wavelength without active mechanical scanning.

\begin{itemize}[nosep]
    \item \textbf{Operational Principle:} Residual atmospheric tip-tilt jitter provides passive position diversity across the photonic lantern input. This inherent motion allows the spatial response function of the PL to be mapped without dynamic beam steering.
    \item \textbf{Required Data:}
    \begin{itemize}[nosep]
        \item Unmodulated PL science spectra (TT mirror held static).
        \item Simultaneous visible or IR focal-plane PSF imaging for real-time spatial jitter monitoring.
        \item Standard off-sky calibration frames (darks, flat-fields, and Neon lamp spectra).
    \end{itemize}
    \item \textbf{Calibration Overhead:} Minimal. No dedicated on-sky calibrator star acquisition is required, as simultaneous focal-plane imaging tracks residual optical alignment variations during observation.
    \item \textbf{Primary Science Regimes:} Sub-$\lambda/D$ structures, young stellar objects (YSOs, tracing accretion and outflow kinematics), Herbig Ae/Be stars, and stellar rotation axis orientations.
\end{itemize}

\subsection{Image Reconstruction Modes}
\label{subsec:image_reconstruction}

For spatial reconstruction across extended fields ($1$--$100~\lambda/D$), FIRST-PL utilizes active spatial modulation via the fast TT mirror to measure the PL spatial response.

\subsubsection{On-Axis Image Reconstruction ($1$--$5~\lambda/D$)}
\label{subsubsec:on_axis_imaging}

\begin{itemize}[nosep]
    \item \textbf{Operational Principle:} Active scanning with the TT mirror modulates the focal-plane beam position across the input face of the photonic lantern. The spatial response function measured on an unresolvable calibrator star is subsequently applied to deconvolve the science target data and reconstruct its spatial intensity distribution.
    \item \textbf{Observing Procedure:}
    \begin{enumerate}[nosep]
        \item Point the telescope to a point-source calibrator star (selected to match or exceed the science target brightness).
        \item Acquire PL calibration data while actively modulating the TT mirror in a predefined modulation pattern.
        \item Slew and acquire the science target.
        \item Acquire PL science spectra using the identical TT modulation pattern.
    \end{enumerate}
    \item \textbf{Calibrator Overheads:} High (requires telescope slews, AO loop acquisition, and modulated data acquisition on both targets).
    \item \textbf{Primary Science Targets:} Close binaries, circumstellar envelopes, and mass-loss structures surrounding evolved stars.
\end{itemize}

\subsubsection{Off-Axis Image Reconstruction (Up to $\sim 100~\lambda/D$ / $2''$)}
\label{subsubsec:off_axis_imaging}

\begin{itemize}[nosep]
    \item \textbf{Operational Principle:} Tailored for high-contrast, wide-separation companions. The bright central star acts as an in-situ calibrator to measure the local PL response matrix prior to acquiring data on the off-axis target.
    \item \textbf{Observing Procedure:}
    \begin{enumerate}[nosep]
        \item Center the telescope on the bright central host star.
        \item Acquire PL calibration data with TT modulation on the host star to calibrate the local response.
        \item Apply a precise target offset to place the off-axis companion onto the PL input core.
        \item Acquire PL science data while executing the TT modulation pattern on the off-axis companion.
    \end{enumerate}
    \item \textbf{Calibrator Overheads:} Low (eliminates target slews and AO re-acquisitions; overheads are restricted to target pointing offsets).
    \item \textbf{Primary Science Targets:} Wide-separation binaries, substellar companions, and exoplanets.
\end{itemize}

\section{INSTRUMENT PERFORMANCE AND ON-SKY COMMISSIONING RESULTS}
\label{sec:commissioning_results}

Following integration on SCExAO, FIRST-PL underwent on-sky commissioning at the Subaru Telescope. In this section, we summarize the measured optical throughput, sensitivity limits, and data reduction methods, showcased alongside representative on-sky results across both observing modes.

\subsection{On-Sky Throughput and Injection Performance}
\label{subsec:throughput}

\begin{figure*}[htbp]
    \centering
    \includegraphics[width=0.6\textwidth]{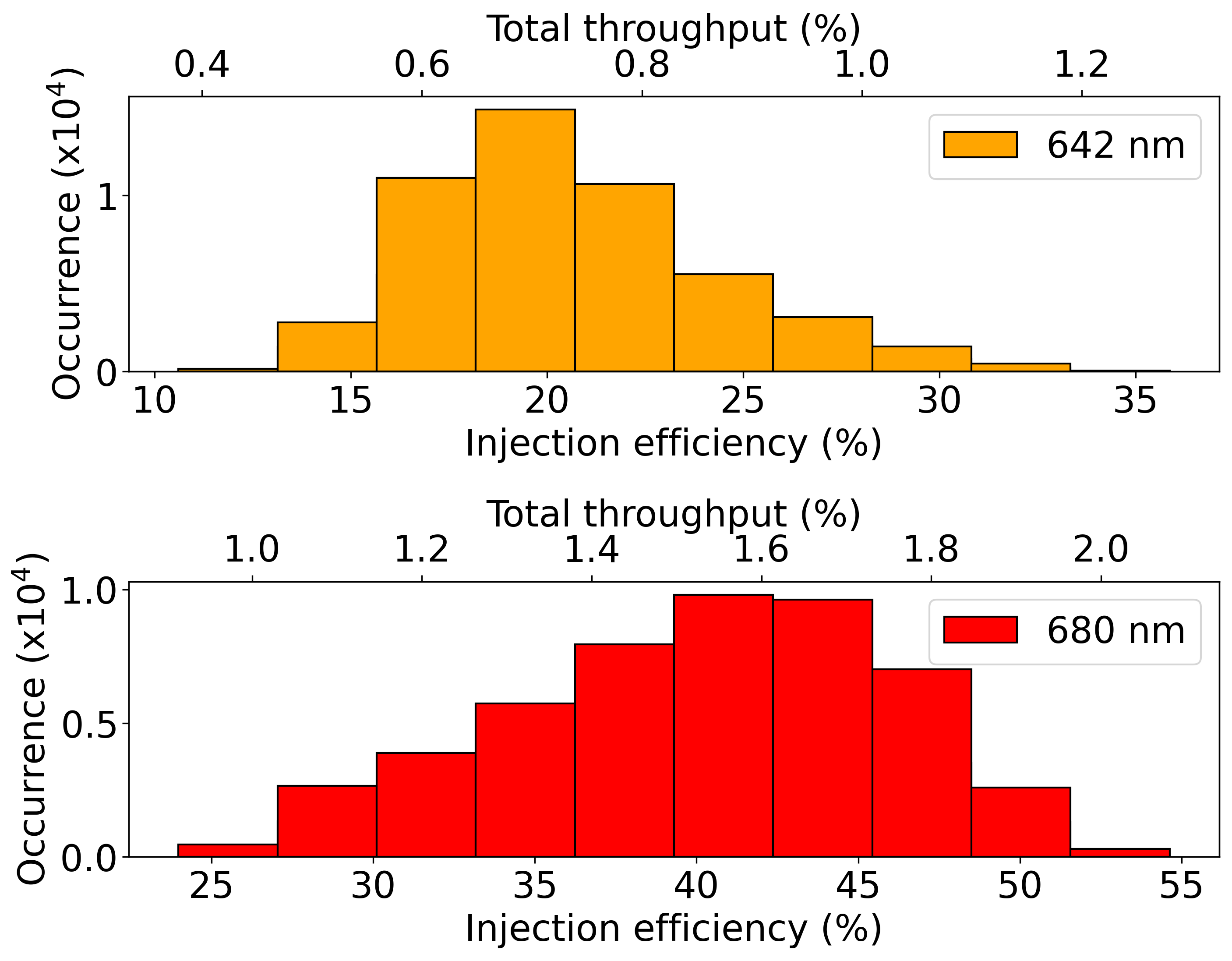}
    \caption{On-sky total throughput and injection efficiency performance at 642 nm (top) and 680 nm (bottom), estimated thanks to the observation of Humu during the S24B observation semester on the Subaru Telescope.}
    \label{fig:injection_eff}
\end{figure*}

Engineering bservations of Humu (Altair, $\alpha$~Aql) during the S24B observing semester evaluated the PL injection efficiency. The results are presented on Figure~\ref{fig:injection_eff} in the form of histograms representing the injection efficiency and total throughput of the instrument (throughput from top of the atmosphere down to the detector) computed on the Humu data at two wavelengths (642nm and 680nm).

At $642\text{~nm}$, the on-sky injection efficiency reached a mean of $21\%$ (peak $36\%$), compared to the laboratory ceiling of $\sim 60\%$. At $680\text{~nm}$, the mean injection efficiency increased to $40\%$ (peak $55\%$), reflecting improved wavefront stability at longer wavelengths. Taking into account losses from telescope reflections, instrument optics, fiber transmission, and detector quantum efficiency (QE), the end-to-end total throughput (top-of-the-atmosphere to detected photons) averages $\sim 1\%$.

\subsection{Sensitivity and Noise Budget}
\label{subsec:sensitivity}

The primary noise sources associated with the science detector include readout noise ($\sigma_{\text{read}} \approx 0.25\text{ e}^-/\text{pixel}$ in standard/slow readout mode) and shot noise from residual stray light background ($I_{\text{bg}} \approx 0.03\text{ e}^-/\text{pixel/s}$). Considering a maximum standard single-frame exposure time of $10\text{ s}$, the total noise per pixel is given by:
\begin{equation}
\sigma_{\text{pixel}} = \sqrt{I_{\text{bg}} \cdot t_{\text{exp}} + \sigma_{\text{read}}^2} = \sqrt{0.3 + 0.25^2} \approx 0.60\text{ e}^-/\text{pixel per } 10\text{ s}.
\end{equation}
Because each spectral trace spans 5 pixels along the direction perpendicular to dispersion, the combined noise per wavelength channel sums quadratically to $\sqrt{5} \times 0.60 \approx 1.35\text{ e}^-$ per $10\text{ s}$.

The science signal corresponds to the number of detected photo-electrons per spectral element, computed assuming a spectral resolution $R \sim 3\,000$, 19 photonic lantern output channels, 3 wavelength channels per resolution element, and an end-to-end efficiency of $1\%$ (combining telescope, SCExAO, instrument optics, and detector quantum efficiency):
\begin{itemize}
    \item \textbf{Bright Reference Target ($R = 0\text{ mag}$):} Produces approximately $1.5 \times 10^6\text{ e}^-$ per spectral element per $10\text{ s}$ frame.
    \item \textbf{Limiting Sensitivity ($R = 12\text{ mag}$):} Defined by a signal-to-noise ratio threshold of $\text{SNR} = 5$ per spectral element, which requires detecting 25 photons ($5\text{ e}^-$ shot noise). For an $R = 12\text{ mag}$ target, the expected signal is $\approx 25\text{ e}^-$ per $10\text{ s}$. At this level, photon shot noise ($5\text{ e}^-$) cleanly dominates over the accumulated detector noise ($1.35\text{ e}^-$).
\end{itemize}

Consequently, the standard limiting sensitivity for FIRST-PL is $R = 12\text{ mag}$. While increasing individual exposure times beyond $10\text{ s}$ could modestly extend this limit, longer frames are generally not recommended during nominal operations to avoid atmospheric decorrelation. 

Furthermore, while $\text{SNR} = 5$ is sufficient for basic image detection, sub-diffraction spectro-astrometry and high-contrast imaging demand substantially higher photon counts. Astrometric precision scales inversely with signal-to-noise ratio ($\sigma_{\text{astro}} \propto 1/\text{SNR}$). Reaching an astrometric accuracy $100\times$ finer than the telescope diffraction limit ($\sim 20\text{ mas}$) requires $\text{SNR} \sim 10^4$, achievable in approximately $30\text{ minutes}$ of total integration time on an $R \approx 5\text{ mag}$ star.

\subsection{Spectro-Astrometry Science Demonstrations}
\label{subsec:sa_results}

Data reduction for spectro-astrometry utilizes real-time focal-plane PSF telemetry from VAMPIRES or NIR channels to track residual spatial jitter and calibrate the spatial response matrix of the 19-port lantern. Detailed theoretical and pipeline descriptions are presented in Kim et al. (2025)~\cite{kim_potential_2024} and Walk et al. (in prep.).

\subsubsection{Decretion Disk of $\beta$ CMi}
\begin{figure}[htbp]
\centering
\includegraphics[width=0.8\textwidth]{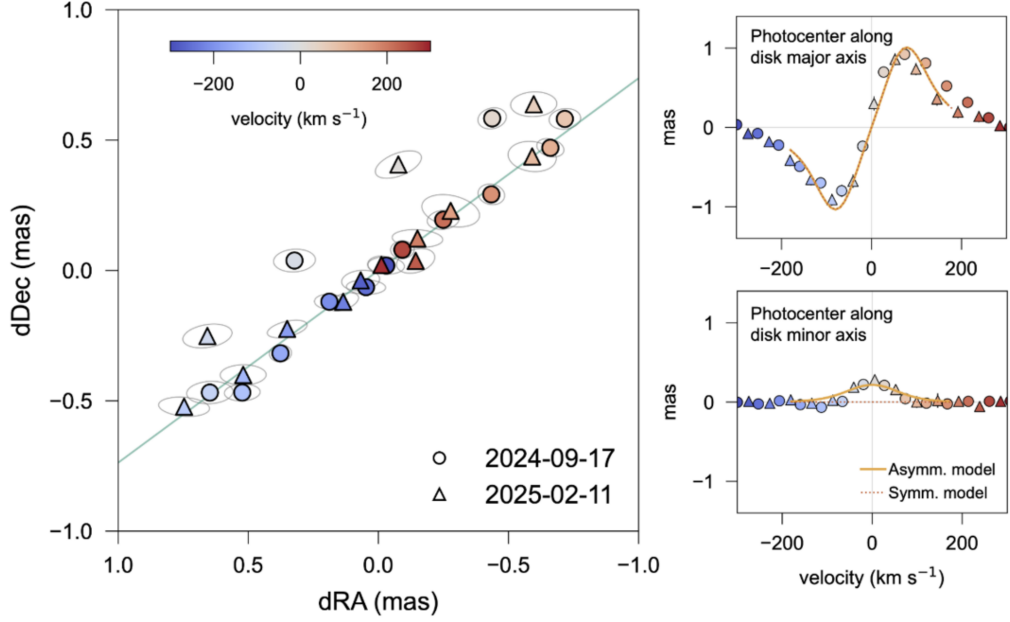}
\caption{Spectro-astrometric measurement of $\beta$ CMi across the H$\alpha$ line using the 19-port spectrograph ($R \sim 3\,000$). Left: Photocenter shift as a function of Doppler velocity. Right: Decomposed spectro-astrometric signatures along the major axis (top) and minor axis (bottom). (Adapted from Kim et al. 2025~\cite{kim_potential_2024}).}
\label{fig:betacmi_sa}
\end{figure}

As reported by Kim et al. (2025)~\cite{kim_potential_2024}, FIRST-PL targeted the classical Be star $\beta$~CMi to demonstrate the capabilities of spatial mode-based imaging on circumstellar environments. Figure~\ref{fig:betacmi_sa} shows the extracted photocenter displacement as a function of Doppler velocity across the H$\alpha$ emission line ($656.3\text{~nm}$). 

The dominant spatial signal detected across the H$\alpha$ line reveals a clear, symmetric photocenter shift along the disk's major axis (position angle of $126^\circ$). This distribution directly traces the Keplerian velocity profile of the gaseous decretion disk at sub-$\lambda/D$ angular scales, in strong agreement with previous studies. 

Furthermore, the high precision of the instrument revealed a subtle minor-axis photocenter shift of $0.27\text{~mas}$. While simple, optically thin, axisymmetric disk models predict zero displacement along the minor axis, this observed shift highlights a distinct near-far side brightness asymmetry driven by opacity effects, where the inner midplane region obscures light emitted behind it.

Achieved from just $\sim 10$ minutes of on-sky observation on an $8.2\text{~m}$ telescope, these measurements reached an unprecedented photocenter precision of $\sim 50\,\mu\text{as}$. The high repeatability across independent epochs (September 2024 and February 2025) confirms that spatial mode-based spectro-astrometry can reliably isolate faint physical disk asymmetries from atmospheric noise.

\subsubsection{Stellar Spin Axis Retrieval of Humu (Altair, $\alpha$ Aql)}
FIRST-PL was further utilized to measure the spin axis orientation of the rapidly rotating star Humu via H$\alpha$ line ($656.46\text{~nm}$) spectro-astrometry~\cite{walk2026first}. Because Humu's angular size is more than five times smaller than the telescope's diffraction limit ($\lambda/D$), traditional direct imaging cannot resolve its stellar disk. 

\begin{figure*}[htbp]
\centering
\includegraphics[width=0.9\textwidth]{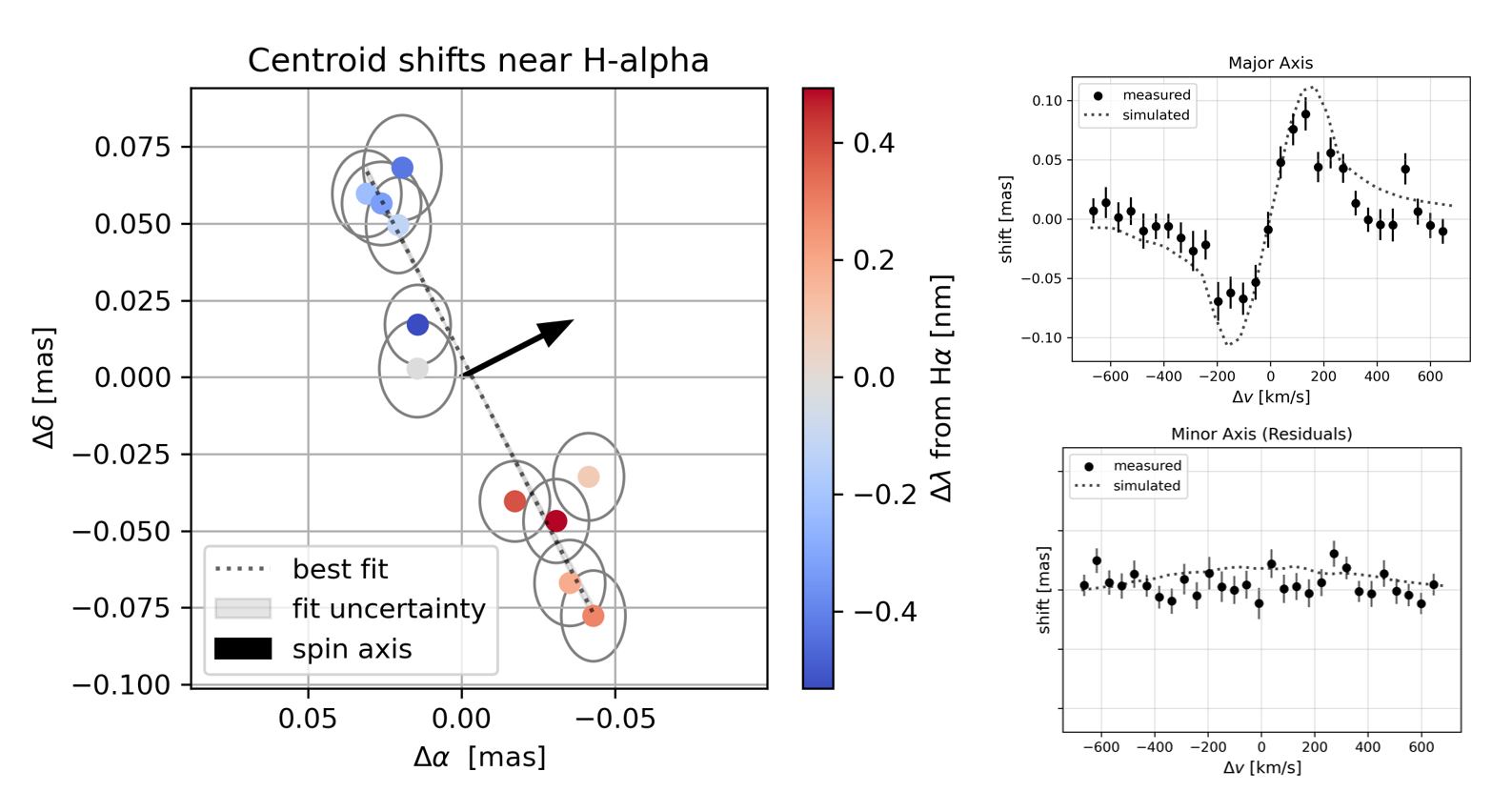}
\caption{Spectro-astrometric measurement of Humu across the H$\alpha$ absorption line. Left: On-sky measured centroid shifts ($\Delta\alpha$, $\Delta\delta$) colored by wavelength offset from H$\alpha$ line center, with $1\sigma$ uncertainty ellipses. The black arrow indicates the derived spin axis position angle. Right: Decomposed spectro-astrometric signatures along the major axis (top) and minor axis (bottom) compared against simulated model signatures (dotted line). (Adapted from Walk et al. 2026~\cite{walk2026first}).}
\label{fig:altair_sa}
\end{figure*}

As shown in Figure~\ref{fig:altair_sa}, using 100 seconds of synchronized, on-sky integration with SCExAO/VAMPIRES, FIRST-PL achieved an average 3$\sigma$ spectro-astrometric precision of $35\,\mu\text{as}$. By tracking the wavelength-dependent centroid shifts across the H$\alpha$ absorption profile, the position angle of Humu's rotation axis was measured to be $-61.8^\circ \pm 2.8^\circ$ (East of North), which agrees within $1\sigma$ with past long-baseline interferometric results from VLTI~\cite{bouchaud2020realistic} and CHARA~\cite{monnier_imaging_2007}.

\begin{figure}[h!]
    \centering
    \includegraphics[width=0.6\textwidth]{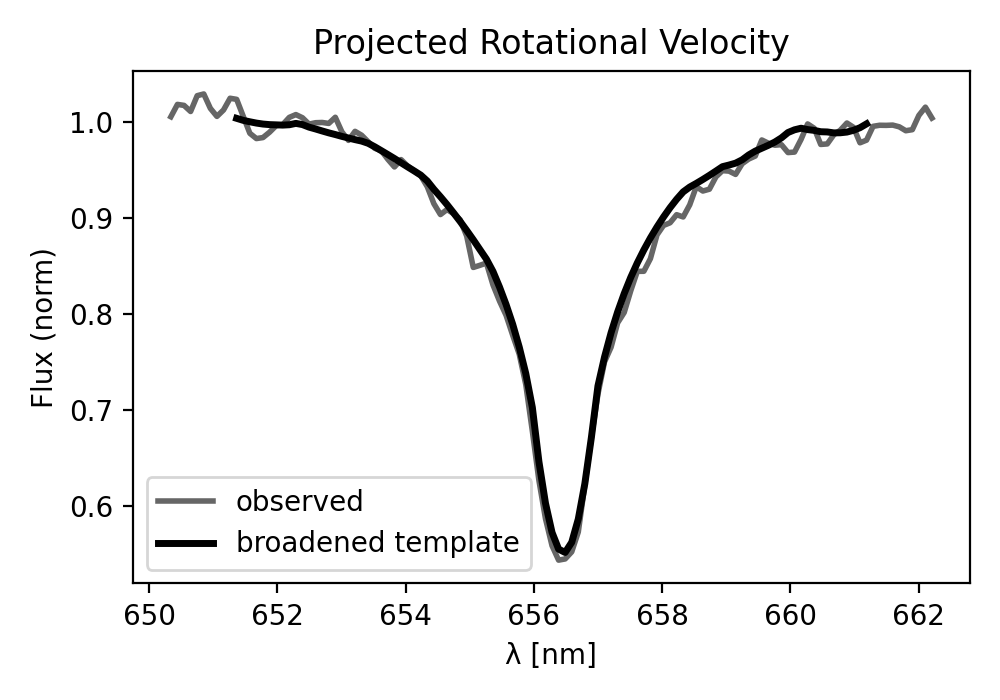}
    \caption{Observed port-summed spectrum of Humu (grey) fitted with the best-fit rotationally broadened synthetic template spectrum ($v \sin i = 220\text{~km/s}$, black)~\cite{walk2026first}.}
\label{fig:altair_vsini}
\end{figure}

Furthermore, as illustrated in Figure~\ref{fig:altair_vsini}, by convolving a PHOENIX synthetic spectrum template with a rotational broadening kernel, a projected rotational velocity of $v \sin i = 220 \pm 16\text{~km/s}$ was extracted. Combining this measurement with literature values for Humu's equatorial radius ($R_{\text{eq}} = 2.0\,R_\odot$) and rotation period ($P = 7.77\text{~h}$) yielded a spin-axis inclination angle of $i = 44.9^\circ \pm 4.1^\circ$. 

These results demonstrate that photonic-lantern-based spectro-astrometry on a single large aperture can reach sub-diffraction spatial scales nearly an order of magnitude more precise than traditional long-slit techniques~\cite{oudmaijer2008, pontoppidan2011}. This single-telescope approach directly competes with the spatial precision of long-baseline optical interferometers—such as CHARA or VLTI/GRAVITY—while drastically bypassing their operational complexity. Rather than requiring multiple telescopes, extensive delay lines, and multi-hour on-sky baseline diversity, photonic-lantern spectro-astrometry recovers comparable sub-diffraction parameters in just tens of seconds of on-sky integration. This offering provides a streamlined, scalable route to constrain host-star spin orientations and infer the orbital geometries of non-transiting exoplanet systems for future facilities like HWO and ELTs.

\subsection{Image Reconstruction and High-Contrast Imaging}
\label{subsec:image_reconstructed_results}

For active beam-modulated datasets, image reconstruction procedures apply deconvolutions using the local PL spatial response matrix calibrated on an unresolvable reference star Details of the data reduction process will be available in Sarrazin et al. (in prep.). 

To demonstrate this mode, FIRST-PL targeted the close binary system HIP 81126 (separation $\sim 70\text{~mas}$) using fast tip-tilt modulation on both the target and an off-axis reference. The resulting image reconstruction will be published in an upcoming paper (Sarrazin et al., in prep.). 

\begin{figure}[htbp]
\centering
\includegraphics[width=0.48\textwidth]{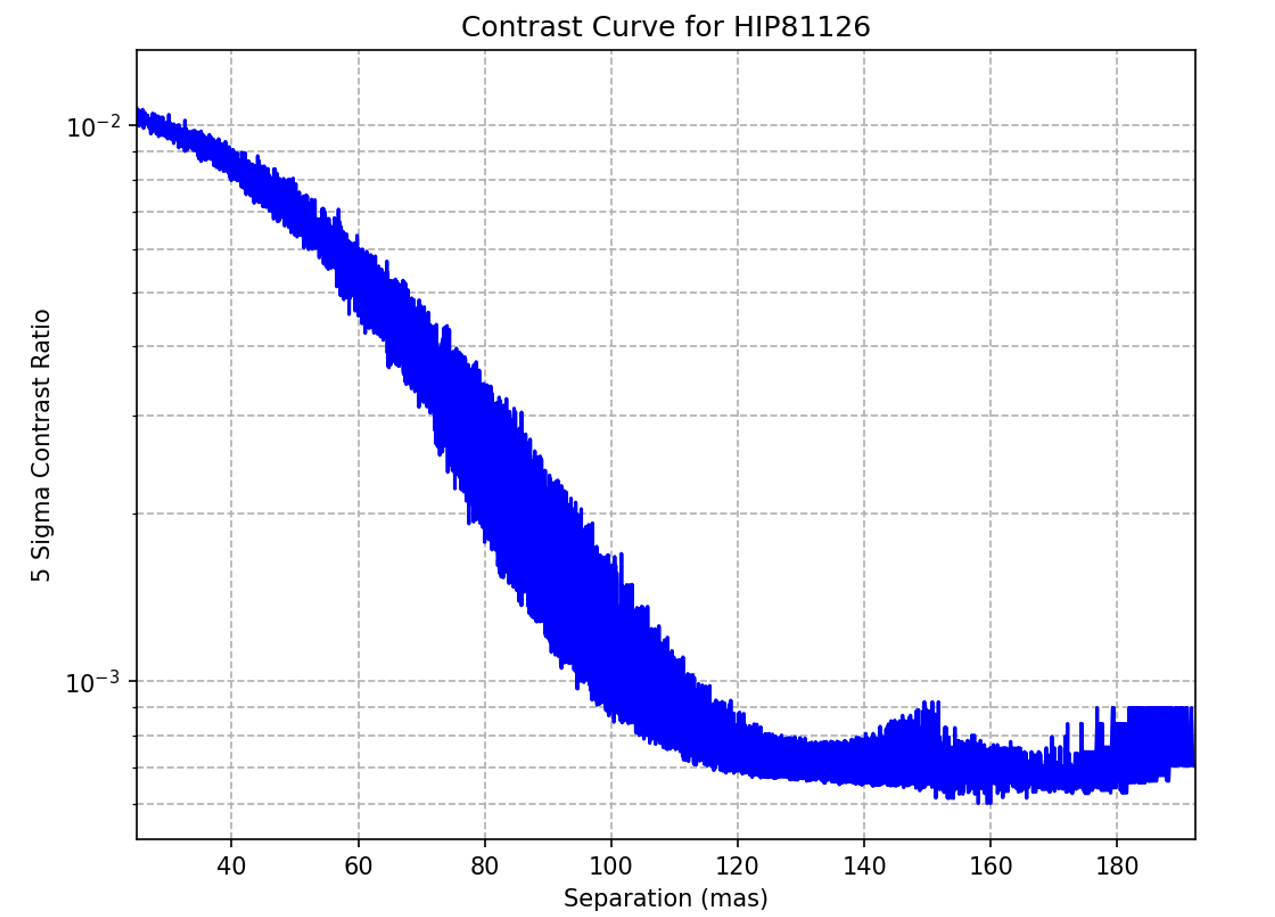}
\caption{$5\sigma$ raw contrast curve for HIP 81126 obtained with FIRST-PL over a $30\text{~second}$ integration time, plotted as a function of angular separation in milliarcseconds.}
\label{fig:hip81126_contrast}
\end{figure}

Here, we present the contrast curve obtained for this target (Figure~\ref{fig:hip81126_contrast}). With a total integration time of only $30\text{~seconds}$, FIRST-PL achieved a $5\sigma$ raw contrast limit of $\sim 10^{-3}$ at $100\text{~mas}$ separation, demonstrating the strong potential of photonic-lantern-based integral field spectroscopy for close-companion detection.

\section{DISCUSSION AND FUTURE OUTLOOK}
\label{sec:discussion}

Following its successful on-sky commissioning and validation, FIRST-PL has transitioned from an experimental setup to an open-community instrument at the Subaru Telescope, officially offered for open-time observations starting in Semester 2026B (S26B). While routine science operations commence, several advanced observational modes and instrument upgrades are currently under active development to further push the boundaries of single-telescope photonic spectro-astrometry and wavefront control.

\subsection{Tip-Tilt Modulated Spectro-Astrometry}
As AO3k+SCExAO performance improves, the natural tip-tilt jitter—and thus the focal-plane spatial diversity—decreases significantly. While beneficial for classical imaging, this reduced spatial sampling complicates spectro-astrometric calibration and required manual PSF dithering during observations. To overcome this limitation and streamline operations, a dedicated acquisition mode using the TT mirror modulation was recently implemented and validated on-sky. This approach dynamically introduces controlled spatial diversity across the photonic lantern aperture. Recent re-analysis of Humu (Altair) data acquired in this mode demonstrates a precision approaching $\sim 10\,\mu\text{as}$ for the star's rotation axis—representing nearly a factor of three improvement over static observations within comparable integration times.

\subsection{High-Resolution Spectro-Astrometric Mode}
While the initial Humu results demonstrate the power of photonic-lantern spectro-astrometry for rapidly rotating stars ($v \sin i \approx 220\text{~km/s}$) at moderate resolving power ($R \sim 3{,}000$), key high-priority science cases require extending this capability to slowly rotating, Sun-like stars ($v \sin i \sim 1$--$5\text{~km/s}$). In particular, determining stellar spin-axis inclinations for known radial-velocity (RV) planet hosts will critically break the $m \sin i$ mass degeneracy and inform target selection for future direct-imaging flagships like HWO and ELTs. Measuring these subtle signatures requires two key enhancements: higher spectral resolution to resolve narrow photospheric absorption lines, and high per-resolution-element SNR to detect microarcsecond-level photocenter displacements.

To achieve this, we are developing a high-resolution ($R \approx 63{,}000$) echelle spectrograph designed to interface directly with the photonic lantern array, covering $620$--$780\text{~nm}$. Developed at the University of Hawaii testbed in collaboration with the Astrobiology Center (ABC) of Japan, this compact architecture utilizes an echelle grating paired with a holographic cross-disperser to format the output of 46 parallel single-mode fiber channels onto a 2D detector. The detailed optical design, mechanical layout, and laboratory performance of this instrument are presented in detail by Hamner et al.~\cite{hamner2026}.

\subsection{Chromatic Wavefront Sensing for Low Wind Effect Mitigation}
Beyond science extraction, the spectrally dispersed modal output of the photonic lantern offers a unique diagnostic capability for real-time wavefront sensing (WFS). A particularly promising application is the mitigation of the Low Wind Effect (LWE)—a severe phase discontinuity across the spider structures of large apertures under low-wind conditions~\cite{milli_low_2018,vievard_focal_2020}. Because LWE phase steps are large ($> 1\,\mu\text{m}$), traditional single-wavelength focal-plane WFS techniques suffer heavily from phase wrapping ambiguity. The spectrally dispersed outputs of the photonic lantern act as a chromatic wavefront sensor, where broad wavelength coverage breaks phase wrapping limits and enables unambiguous phase-step retrieval across telescope spiders. Integrating this photonic lantern telemetry into the SCExAO real-time control loop promises a robust pathway to solve the LWE on current and ELT-scale facilities.

\section{CONCLUSION}
\label{sec:conclusion}

FIRST-PL provides a novel, high-throughput solution for sub-diffraction spectro-imaging on 8-meter class telescopes. Directly downstream of the double-stage AO system (AO3K and SCExAO), the instrument leverages extreme wavefront stability to feed a 19-port visible photonic lantern coupled to a medium-resolution ($R \sim 3{,}000$) spectrograph optimized for visible wavelengths ( $620 - 780\text{ nm}$). Its versatile operating modes support both narrow-field IFU-style spatial reconstruction and precision spectro-astrometry far below the telescope's classical diffraction limit.

On-sky commissioning has successfully validated the instrument's capabilities:
\begin{itemize}[nosep]
    \item \textbf{Circumstellar Disk Kinematics ($\beta$ CMi):} Recovered the sub-diffraction Keplerian disk velocity map and detected a subtle $0.27\text{~mas}$ minor-axis brightness asymmetry, revealing previously unobserved opacity structure.
    \item \textbf{Stellar Spin-Axis Orientations (Humu):} Reached double-digit microarcsecond precision ($\sim 35\,\mu\text{as}$) in just $100\text{~seconds}$ of total integration time, obtaining spin-axis measurements at spatial scales far surpassing traditional slit spectro-astrometry and competing directly with long-baseline optical interferometry.
    \item \textbf{Image Reconstruction (HIP 81126):} Demonstrated IFU-style companion detection by achieving a $5\sigma$ raw contrast limit of $\sim 10^{-3}$ at an angular separation of $100\text{~mas}$ in only $30\text{~seconds}$ of integration.
\end{itemize}

These results highlight a very promising pathway for single-aperture sub-diffraction astronomy. FIRST-PL is now officially open for community science at the Subaru Telescope starting in Semester 2026B (S26B). Operational enhancements—including active tip-tilt mirror modulation—are already yielding even higher spectro-astrometric precisions ($\sim 10\,\mu\text{as}$) while simplifying operations. Looking forward, an upcoming high-resolution mode ($R \sim 63{,}000$) and the exploitation of the photonic lantern as a chromatic wavefront sensor to mitigate the Low Wind Effect (LWE) will further broaden FIRST-PL's scientific impact, providing key empirical inputs for future direct-imaging programs on HWO and ELTs.

\acknowledgments 
The development of FIRST-PL was supported by the French National Research Agency (ANR-21-CE31-0005; ANR-21-CE31-0017; ANR-22-EXOR-0005). The development of SCExAO is supported by the Japan Society for the Promotion of Science (Grant-in-Aid for Research \#23340051, \#26220704, \#23103002, \#19H00703, \#19H00695 and \#21H04998), the Subaru Telescope, the National Astronomical Observatory of Japan, the Astrobiology Center of the National Institutes of Natural Sciences, Japan, the Mt Cuba Foundation and the Heising-Simons Foundation. The development of the CACAO software is supported by the National Science Foundation under award 2410616. The authors wish to recognize and acknowledge the very significant cultural role and reverence that the summit of Maunakea has always had within the indigenous Hawaiian community, and are most fortunate to have the opportunity to conduct observations from this mountain.

\bibliography{report.bib} 

\begin{thebibliography}{10}

\bibitem{snellen10}
{Snellen}, I. A.~G., {de Kok}, R.~J., {de Mooij}, E. J.~W., and {Albrecht}, S.,
  ``{The orbital motion, absolute mass and high-altitude winds of exoplanet
  HD209458b},'' {\em Nature}~{\bf 465},  1049--1051 (June 2010).

\bibitem{brogi12}
{Brogi}, M., {Snellen}, I. A.~G., {de Kok}, R.~J., {Albrecht}, S., {Birkby},
  J., and {de Mooij}, E. J.~W., ``{The signature of orbital motion from the
  dayside of the planet {\ensuremath{\tau}} Bo{\"o}tis b},'' {\em Nature}~{\bf
  486},  502--504 (June 2012).

\bibitem{currie2023ppvii}
{Currie}, T., {Biller}, B., {Lagrange}, A., {Marois}, C., {Guyon}, O.,
  {Nielsen}, E.~L., {Bonnefoy}, M., and {De Rosa}, R.~J., ``{Direct Imaging and
  Spectroscopy of Extrasolar Planets},'' in [{\em Protostars and Planets
  VII}{\nolinebreak\hspace{0.1em}]},  {Inutsuka}, S., {Aikawa}, Y., {Muto}, T.,
  {Tomida}, K., and {Tamura}, M., eds., {\em Astronomical Society of the
  Pacific Conference Series} {\bf 534},  799 (July 2023).

\bibitem{bryan_constraints_2017}
Bryan, M.~L., Benneke, B., Knutson, H.~A., Batygin, K., and Bowler, B.~P.,
  ``Constraints on the spin evolution of young planetary-mass companions,''
  {\em Nature Astronomy}~{\bf 2},  138--144 (Dec. 2017).

\bibitem{bryan_obliquity_2020}
Bryan, M.~L., Chiang, E., Bowler, B.~P., Morley, C.~V., Millholland, S., Blunt,
  S., Ashok, K.~B., Nielsen, E., Ngo, H., Mawet, D., and Knutson, H.~A.,
  ``Obliquity {Constraints} on an {Extrasolar} {Planetary}-mass {Companion},''
  {\em The Astronomical Journal}~{\bf 159},  181 (Apr. 2020).

\bibitem{lissauer1993planet}
Lissauer, J.~J., ``Planet formation,'' {\em In: Annual review of astronomy and
  astrophysics. Vol. 31 (A94-12726 02-90), p. 129-174.}~{\bf 31},  129--174
  (1993).

\bibitem{kotani2020reach}
Kotani, T., Kawahara, H., Ishizuka, M., Jovanovic, N., Guyon, O., Vievard, S.,
  Lozi, J., Sahoo, A., Yoneta, K., and Tamura, M., ``The reach project:
  combining extremely high-contrast and high spectral resolution at the subaru
  telescope,'' in [{\em Adaptive Optics Systems
  VII}{\nolinebreak\hspace{0.1em}]},   {\bf 11448},  1144878, International
  Society for Optics and Photonics (2020).

\bibitem{delorme2021keck}
Delorme, J.-R., Jovanovic, N., Echeverri, D., Mawet, D., Kent~Wallace, J.,
  Bartos, R.~D., Cetre, S., Wizinowich, P., Ragland, S., Lilley, S., et~al.,
  ``Keck planet imager and characterizer: a dedicated single-mode fiber
  injection unit for high-resolution exoplanet spectroscopy,'' {\em Journal of
  Astronomical Telescopes, Instruments, and Systems}~{\bf 7}(3),
  035006--035006 (2021).

\bibitem{martinod2021scalable}
Martinod, M.-A., Norris, B., Tuthill, P., Lagadec, T., Jovanovic, N.,
  Cvetojevic, N., Gross, S., Arriola, A., Gretzinger, T., Withford, M.~J.,
  et~al., ``Scalable photonic-based nulling interferometry with the dispersed
  multi-baseline glint instrument,'' {\em Nature communications}~{\bf 12}(1),
  2465 (2021).

\bibitem{vievard2023single}
Vievard, S., Huby, E., Lacour, S., Guyon, O., Cvetojevic, N., Jovanovic, N.,
  Lozi, J., Barjot, K., Deo, V., Duch{\^e}ne, G., et~al., ``Single-aperture
  spectro-interferometry in the visible at the subaru telescope with first:
  First on-sky demonstration on keho ‘oea ($\alpha$ lyrae) and hokulei
  ($\alpha$ aurigae),'' {\em Astronomy \& Astrophysics}~{\bf 677},  A84 (2023).

\bibitem{vigan2024first}
Vigan, A., El~Morsy, M., Lopez, M., Otten, G., Garcia, J., Costes, J.,
  Muslimov, E., Viret, A., Charles, Y., Zins, G., et~al., ``First light of
  vlt/hirise: High-resolution spectroscopy of young giant exoplanets,'' {\em
  Astronomy \& Astrophysics}~{\bf 682},  A16 (2024).

\bibitem{mawet2024fiber}
Mawet, D., Fitzgerald, M.~P., Konopacky, Q., Jovanovic, N., Baker, A.,
  Andersen, D., Artigau, E., Bailey~III, J.~I., Beichman, C., Benneke, B.,
  et~al., ``Fiber-fed high-resolution infrared spectroscopy at the diffraction
  limit with keck-hispec and tmt-modhis: status update,'' in [{\em Ground-based
  and Airborne Instrumentation for Astronomy X}{\nolinebreak\hspace{0.1em}]},
  {\bf 13096},  227--237, SPIE (2024).

\bibitem{haniff1987first}
Haniff, C.~A., Mackay, C., Titterington, D.~J., Sivia, D., Baldwin, J., and
  Warner, P., ``The first images from optical aperture synthesis,'' {\em
  Nature}~{\bf 328}(6132),  694--696 (1987).

\bibitem{gauchet2016sparse}
Gauchet, L., Lacour, S., Lagrange, A.-M., Ehrenreich, D., Bonnefoy, M., Girard,
  J., and Boccaletti, A., ``Sparse aperture masking at the vlt-ii. detection
  limits for the eight debris disks stars $\beta$ pic, au mic, 49 cet, $\eta$
  tel, fomalhaut, g lup, hd 181327 and hr 8799,'' {\em Astronomy \&
  Astrophysics}~{\bf 595},  A31 (2016).

\bibitem{perrin_high_2006}
Perrin, G., Lacour, S., Woillez, J., and Thiebaut, E., ``High dynamic range
  imaging by pupil single-mode filtering and remapping,'' {\em Monthly Notices
  of the Royal Astronomical Society}~{\bf 373},  747--751 (Dec. 2006).

\bibitem{huby_first_2013}
Huby, E., Duchêne, G., Marchis, F., Lacour, S., Perrin, G., Kotani, T.,
  Choquet, A, Gates, E.~L., Lai, O., and Allard, F., ``{FIRST}, a fibered
  aperture masking instrument: {II}. {Spectroscopy} of the {Capella} binary
  system at the diffraction limit,'' {\em Astronomy \& Astrophysics}~{\bf 560},
   A113 (Dec. 2013).

\bibitem{jovanovic_subaru_2015}
Jovanovic, N., Martinache, F., Guyon, O., Clergeon, C., Singh, G., Kudo, T.,
  Garrel, V., Newman, K., Doughty, D., Lozi, J., Males, J., Minowa, Y., Hayano,
  Y., Takato, N., Morino, J., Kuhn, J., Serabyn, E., Norris, B., Tuthill, P.,
  Schworer, G., Stewart, P., Close, L., Huby, E., Perrin, G., Lacour, S.,
  Gauchet, L., Vievard, S., Murakami, N., Oshiyama, F., Baba, N., Matsuo, T.,
  Nishikawa, J., Tamura, M., Lai, O., Marchis, F., Duchene, G., Kotani, T., and
  Woillez, J., ``The {Subaru} {Coronagraphic} {Extreme} {Adaptive} {Optics}
  {System}: {Enabling} {High}-{Contrast} {Imaging} on {Solar}-{System}
  {Scales},'' {\em Publications of the Astronomical Society of the
  Pacific}~{\bf 127},  890--910 (Sept. 2015).

\bibitem{Birks:15}
Birks, T.~A., Gris-S\'{a}nchez, I., Yerolatsitis, S., Leon-Saval, S.~G., and
  Thomson, R.~R., ``The photonic lantern,'' {\em Adv. Opt. Photon.}~{\bf 7},
  107--167 (Jun 2015).

\bibitem{El_Morsy_2022}
El~Morsy, M., Vigan, A., Lopez, M., Otten, G. P. P.~L., Choquet, E., Madec, F.,
  Costille, A., Sauvage, J.-F., Dohlen, K., Muslimov, E., Pourcelot, R.,
  Floriot, J., Benedetti, J.-A., Blanchard, P., Balard, P., and Murray, G.,
  ``Validation of strategies for coupling exoplanet psfs into single-mode
  fibres for high-dispersion coronagraphy,'' { Astronomy and;
  Astrophysics}~{\bf 667},  A171 (Nov. 2022).

\bibitem{jovanovic_efficient_2017}
Jovanovic, N., Schwab, C., Guyon, O., Lozi, J., Cvetojevic, N., Martinache, F.,
  Leon-Saval, S., Norris, B., Gross, S., Doughty, D., Currie, T., and Takato,
  N., ``Efficient injection from large telescopes into single-mode fibres:
  {Enabling} the era of ultra-precision astronomy,'' {\em Astronomy \&
  Astrophysics}~{\bf 604},  A122 (Aug. 2017).

\bibitem{tamura2016prime}
Tamura, N., Takato, N., Shimono, A., Moritani, Y., Yabe, K., Ishizuka, Y.,
  Ueda, A., Kamata, Y., Aghazarian, H., Arnouts, S., et~al., ``Prime focus
  spectrograph (pfs) for the subaru telescope: overview, recent progress, and
  future perspectives,'' {\em Ground-based and airborne instrumentation for
  astronomy VI}~{\bf 9908},  456--472 (2016).

\bibitem{kim_potential_2024}
Kim, Y.~J., Fitzgerald, M.~P., Lin, J., Xin, Y., Levinstein, D., Sallum, S.,
  Jovanovic, N., and Leon-Saval, S., ``On the {Potential} of
  {Spectroastrometry} with {Photonic} {Lanterns},'' (Sept. 2024).
\newblock arXiv:2409.09120 [astro-ph].

\bibitem{norris2020all}
Norris, B.~R., Wei, J., Betters, C.~H., Wong, A., and Leon-Saval, S.~G., ``An
  all-photonic focal-plane wavefront sensor,'' {\em Nature Communications}~{\bf
  11}(1),  5335 (2020).

\bibitem{norris_all-photonic_2020}
Norris, B. R.~M., Wei, J., Betters, C.~H., Wong, A., and Leon-Saval, S.~G.,
  ``An all-photonic focal-plane wavefront sensor,'' {\em Nature
  Communications}~{\bf 11},  5335 (Oct. 2020).
\newblock arXiv:2003.05158 [astro-ph].

\bibitem{vievard_spectroscopy_2024}
Vievard, S., Lallement, M., Leon-Saval, S., Guyon, O., Jovanovic, N., Huby, E.,
  Lacour, S., Lozi, J., Deo, V., Ahn, K., Lucas, M., Sallum, S., Norris, B.,
  Betters, C., Amezcua-Correa, R., Yerolatsitis, S., Fitzgerald, M.~P., Lin,
  J., Kim, Y.~J., Gatkine, P., Kotani, T., Tamura, M., Currie, T., Kenchington,
  H.-D., Martin, G., and Perrin, G., ``Spectroscopy using a visible photonic
  lantern at the {Subaru} {Telescope}: {Laboratory} characterization and the
  first on-sky demonstration on {Ikiiki} ( alpha {Leo}) and {Aua} (alpha {Ori}),'' { Astronomy and Astrophysics}~{ 691},  A140 (Nov.
  2024).

\bibitem{lozi2024ao3k}
Lozi, J., Ahn, K., Blue, H., Chun, A., Clergeon, C., Deo, V., Guyon, O.,
  Hattori, T., Minowa, Y., Nishiyama, S., et~al., ``Ao3k at subaru: first
  on-sky results of the facility extreme-ao,'' in [{\em Adaptive Optics Systems
  IX}{\nolinebreak\hspace{0.1em}]},   {\bf 13097},  1309703, SPIE (2024).

\bibitem{Lozi_2019}
Lozi, J., Jovanovic, N., Guyon, O., Chun, M., Jacobson, S., Goebel, S., and
  Martinache, F., ``Visible and near-infrared laboratory demonstration of a
  simplified pyramid wavefront sensor,'' {\em Publications of the Astronomical
  Society of the Pacific}~{\bf 131},  044503 (mar 2019).

\bibitem{lucas2024visible}
Lucas, M., Norris, B., Guyon, O., Bottom, M., Deo, V., Vievard, S., Lozi, J.,
  Ahn, K., Ashcraft, J., Currie, T., et~al., ``Visible-light high-contrast
  imaging and polarimetry with scexao/vampires,'' {\em Publications of the
  Astronomical Society of the Pacific}~{\bf 136}(11),  114504 (2024).

\bibitem{walk2026first}
Walk, A., Vievard, S., Guyon, O., Kim, Y.~J., Huby, E., Lacour, S., Lallement,
  M., Nowak, M., Sarrazin, J., Lozi, J., Deo, V., Lucas, M., Fitzgerald, M.,
  Leon-Saval, S., Norris, B., Sallum, S., Currie, T., Tamura, M., and Kotani,
  T., ``{FIRST-PL}: on-sky retrieval of {Humu's} ({Altair}) spin axis using a
  photonic lantern on {Subaru} telescope,'' in [{\em Advances in Optical and
  Mechanical Technologies for Telescopes and Instrumentation
  VII}{\nolinebreak\hspace{0.1em}]},  {\em Proc. SPIE} {\bf 14154},  14154--90,
  SPIE (2026).
\newblock Conference held 5--10 July 2026; presented 9 July 2026.

\bibitem{bouchaud2020realistic}
Bouchaud, K., Domiciano~de Souza, A., Rieutord, M., Reese, D., and Kervella,
  P., ``A realistic two-dimensional model of altair,'' {\em Astronomy \&
  Astrophysics}~{\bf 633},  A78 (2020).

\bibitem{monnier_imaging_2007}
Monnier, J.~D., Zhao, M., Pedretti, E., Thureau, N., Ireland, M., Muirhead, P.,
  Berger, J.~P., Millan-Gabet, R., Van~Belle, G., Brummelaar, T.~t., McAlister,
  H., Ridgway, S., Turner, N., Sturmann, L., Sturmann, J., and Berger, D.,
  ``Imaging the {Surface} of {Altair},'' (2007).
\newblock Version Number: 2.

\bibitem{oudmaijer2008}
Oudmaijer, R.~D., Parr, A., Baines, D., and Porter, J., ``Sub-milliarcsecond
  precision spectro-astrometry of be stars,'' {\em Astronomy \&
  Astrophysics}~{\bf 489}(2),  627--631 (2008).

\bibitem{pontoppidan2011}
Pontoppidan, K.~M., Blake, G.~A., and Smette, A., ``The structure and dynamics
  of molecular gas in planet-forming zones: a crires spectro-astrometric
  survey,'' {\em The Astrophysical Journal}~{\bf 733}(2),  84 (2011).

\bibitem{hamner2026}
Hamner, C., Vievard, S., Gilbert, A., Walk, A., Huby, E., Lacour, S., Martin,
  G., Hirano, T., Tamura, M., Kotani, T., and Guyon, O., ``High resolution
  echelle spectrometer integrating 19 channel photonic lantern,'' in [{\em
  Advances in Optical and Mechanical Technologies for Telescopes and
  Instrumentation VII}{\nolinebreak\hspace{0.1em}]},  {\em Proc. SPIE} {\bf
  14148},  14148--97, SPIE (2026).
\newblock Conference held 5--10 July 2026; presented 9 July 2026.

\bibitem{milli_low_2018}
Milli, J., Mouillet, D., Beuzit, J.-L., Sauvage, J.-F., Fusco, T., Bourget, P.,
  Kasper, M., Tristam, K., Reyes, C., Girard, J.~H., Telle, A., Pannetier, C.,
  Cantalloube, F., Wahhaj, Z., Mawet, D., Vigan, A., and N'Diaye, M., ``Low
  wind effect on {VLT}/{SPHERE}: impact, mitigation strategy, and results,'' in
  [{\em Adaptive {Optics} {Systems} {VI}}{\nolinebreak\hspace{0.1em}]},
  Schmidt, D., Schreiber, L., and Close, L.~M., eds.,  83, SPIE, Austin, United
  States (July 2018).

\bibitem{vievard_focal_2020}
Vievard, S., Bos, S.~P., Cassaing, F., Currie, T., Deo, V., Guyon, O.,
  Jovanovic, N., Keller, C., Lamb, M., Lopez, C., Lozi, J., Martinache, F.,
  Miller, K., Montmerle-Bonnefois, A., Mugnier, L.~M., N'Diaye, M., Norris, B.,
  Sahoo, A., Sauvage, J.-F., Skaf, N., Snik, F., Wilby, M.~J., and Wong, A.,
  ``Focal {Plane} {Wavefront} {Sensing} on {SUBARU}/{SCExAO},'' (Dec. 2020).
\newblock arXiv:2012.12417 [astro-ph].

\end{thebibliography}
\bibliographystyle{spiebib} 
\end{document}